\documentclass[preprint,prd,showpacs]{revtex4} 
 
\usepackage{graphicx}
\usepackage{amsmath}
\usepackage{amsfonts}
\usepackage{amssymb}

\begin{document}

\title{The Electromagnetic Field Stress Tensor between Dielectric Half-Spaces}
\author{V. Sopova}
 \email{vasilka.sopova@tufts.edu}
\author{L. H. Ford}
  \email{ford@cosmos.phy.tufts.edu}
\affiliation{Institute of Cosmology \\ 
Department of Physics and Astronomy\\
 Tufts University, Medford, Massachusetts 02155}

\begin{abstract}
The stress tensor for the quantized electromagnetic field is calculated
in the region between a pair of dispersive, dielectric half-spaces.
This generalizes the stress tensor for the Casimir energy to the case
where the boundaries have finite reflectivity. We also include the effects of 
finite temperature. This allows us to discuss the circumstances under
which the weak energy condition and the null energy condition can be
violated in the presence of finite reflectivity and finite temperature.
\end{abstract}


\maketitle

\baselineskip=12pt

\section{Introduction}

The Casimir force~\cite{Casimir} between a pair of parallel, perfectly
 reflecting plates
is remarkable prediction of quantum electrodynamics which has been confirmed
by experiment~\cite{Sparnaay,Lamoreaux,Mohideen,Mohideen2,Chan,Bressi}. 
It also has
some implications for the semiclassical theory of gravity, as the stress
tensor of the Casimir energy violates the weak energy condition. Simply from
Casimir's result for the force per unit area, one can construct the entire
stress tensor, using conservation, tracelessness and symmetry 
arguments~\cite{Brown,DeWitt}. The result is
\begin{equation}
T_{\mu\nu} =
\begin{pmatrix} T_{00} & 0 & 0 & 0 \\
                 0 & T_{xx} & 0 & 0 \\
                 0 & 0 & T_{yy} & 0 \\
                 0 & 0 & 0 & T_{zz}
\end{pmatrix}
= \frac{\pi^2}{720\,a^4}\;
\begin{pmatrix} -1 & 0 & 0 & 0 \\
                 0 & 1 & 0 & 0 \\
                 0 & 0 & 1 & 0 \\
                 0 & 0 & 0 & -3
\end{pmatrix}
\end{equation}
Here the plates are separated by a distance $a$ in the $z$-direction, 
and units where $\hbar = c =1$ are used. Here $T_{\mu\nu}$ is understood to be
a renormalized expectation value of the quantum stress tensor operator.

Because the local energy density is negative, the weak energy condition
is violated. In addition, the null energy condition, 
$T_{\mu\nu} k^\mu k^\nu \geq 0$ for all null vectors $k_\mu$ is violated
as well, except for the case where $k_\mu$ is parallel to the plates, in which
case $T_{\mu\nu} k^\mu k^\nu = 0$. The null energy condition is the
condition for gravity to locally focus a bundle of null rays. 
The average null energy condition,
\begin{equation}
\int d\lambda \, T_{\mu\nu} k^\mu k^\nu \geq 0 \, , \label{eq:anec}
\end{equation}
along a complete null geodesic, is more difficult to violate~\cite{GOS}. 
Null rays which
are not parallel to the plates eventually intersect and pass through the 
plates. The integral in Eq.~(\ref{eq:anec}) then gets a contribution from the
matter composing the plates. The extent to which quantum fields could
violate the average null energy condition is of interest in several aspects
of gravity theory. Its violation, for example, is essential to construct
traversable wormholes~\cite{Visser}.

Some authors~\cite{Lang,lam} have suggested that when the assumption of 
perfect conductivity is removed, the negative energy density might disappear.
However, in a previous paper~\cite{Energy}, we calculated the energy density
between two half-spaces filled with dispersive material, and showed that the
energy density in the center can still be negative. However, the energy density
is no longer constant in the region between the interfaces, and diverges 
positively
at the boundaries. This divergence, which arises despite a dispersive 
dielectric 
function which approaches one at high frequency, can be attributed to the
assumption of a sharp boundary between the dielectric and vacuum regions.

It may come as a surprize that the energy density is finite between
perfectly reflecting plates, but diverges near plates of finite reflectivity.
However, in the perfectly reflecting case there is a cancellation between
two divergent terms. Both the mean squared electric field, 
$\langle E^2 \rangle$, and  mean squared magnetic field, 
$\langle B^2 \rangle$, diverge but the energy density is finite. When the
plates have finite reflectivity, both $\langle E^2 \rangle$ and 
$\langle B^2 \rangle$ diverge less rapidly than in the perfectly reflecting 
case, but the cancellation is upset, so that the energy density also
diverges near the plates. There are more complicated geometries where
similar cancellation can occur. In the interior of a wedge with perfecty
reflecting walls, the energy density diverges near the corner, but is finite
if one approaches either wall away from the corner~\cite{NLS}. Again there
must be a cancellation which would be upset if the wall had finite 
reflectivity. 

In the present paper, we will extend the results of Ref.~\cite{Energy} to
study the pressure components of the stress tensor and the effects of finite 
temperature. As before, our calculations are based on a formalism developed
by Schwinger, DeRaad, and Milton\cite{Schwinger}. Lorentz-Heaviside units with
$\hbar = c =1$ will be used.

\section{The Electromagnetic Field Between Dielectric Slabs-Corrections at
Finite Temperature}

\noindent We consider the electromagnetic field stress tensor in the vacuum
region of width $a$\ between two dielectric half-spaces whose dielectric
function is described by the plasma model:%

\begin{gather}
0<z<a:\,\,\epsilon(z)=1,\label{Plasma1}\\
z<0\,\ \mathrm{and\,}\,z>a:\,\epsilon(z)\equiv\epsilon=1-\frac{\omega_{p}^{2}%
}{\omega^{2}}, \label{Plasma2}%
\end{gather}
where $\omega_{p}$ is the plasma frequency. The finite mean squared electric
field in the vacuum region, at zero temperature, can be expressed as an
integral over imaginary frequency $\zeta$ \cite{Energy}

\begin{multline}
\left\langle E^{2}\right\rangle =\frac{1}{2\pi^{2}}\int_{0}^{\infty}d\zeta
\int_{0}^{\infty}dk\frac{\,k}{\kappa}\left\{  \zeta^{2}\left(  \frac{r^{2}%
}{r^{2}-e^{2\kappa a}}+\frac{r^{\prime}{}^{2}}{r^{\prime}{}^{2}-e^{2\kappa a}%
}\right)  +\right. \\
\left.  +\left[  -\zeta^{2}\frac{r}{1-r^{2}e^{-2\kappa a}}+(2k^{2}+\zeta
^{2})\frac{r^{\prime}}{1-r^{\prime}{}^{2}e^{-2\kappa a}}\right]  e^{-\kappa
a}\cosh\left[  \kappa(2z-a)\right]  \right\}\, ,
\end{multline}
where the reflection coefficients for S and P polarizations, respectively,
 are given by
\begin{align}
r  &  =\frac{\kappa-\kappa_{1}}{\kappa+\kappa_{1}}\\
r^{\prime}  &  =\frac{\kappa\epsilon-\kappa_{1}}{\kappa\epsilon+\kappa_{1}}\,.
\end{align}
 The quantities $\kappa$ and $\kappa_{1}$ are defined as
$\kappa^{2}=k^{2}+\zeta^{2}$, and $\kappa_{1}^{2}=k^{2}+\zeta^{2}\epsilon$.
The expression for the mean squared magnetic field $\left\langle
B^{2}\right\rangle$ is obtained from that for $\left\langle E^{2}\right\rangle$
by interchanging the coefficients $r$ and $r^{\prime}$.
The energy density in the vacuum region, $U=  T_{00}  =
\left(  \left\langle
E^{2}\right\rangle +\left\langle B^{2}\right\rangle \right)  /2$ is then%

\begin{eqnarray}
U=\frac{1}{2\pi^{2}}\int_{0}^{\infty}d\zeta\int_{0}^{\infty}%
dk\frac{\,k}{\kappa}\left\{  \zeta^{2}\left(  \frac{r^{2}}{r^{2}-e^{2\kappa
a}}+\frac{r^{\prime}{}^{2}}{r^{\prime}{}^{2}-e^{2\kappa a}}\right)  +\right.
\nonumber  \\
\left.  +k^{2}\left(  \frac{r}{1-r^{2}e^{-2\kappa a}}+\frac{r^{\prime}%
}{1-r^{\prime}{}^{2}e^{-2\kappa a}}\right)  e^{-\kappa a}\cosh\left[
\kappa(2z-a)\right]  \right\} \label{eq:U} .
\end{eqnarray}
As discussed in \cite{Energy}, $U$ is position dependent: it has a minimum at
the center of the vacuum region and diverges at the interfaces. The overall
sign of $U$ at its minimum depends on the choice of $a$ and $\omega_{p}$. As
the product $\omega_{p}a$ grows, $U$ at the midpoint decreases, becoming
negative for $\omega_{p}a\approx100$ (see Figure 2, dash-dot line). It is of
interest to examine the effects due to finite temperature upon the energy
density and see when its sign can still be negative when the temperature is 
not zero.

For this purpose, we write Eq.~(\ref{eq:U}) as a Fourier series instead 
of an integral on $\zeta$ \cite{Schwinger}:%

\begin{multline}
\label{Uct}U_{C}=\frac{1}{\pi\beta}\sum_{n=0}^{\infty}{}^{\prime}\int_{0}^{\infty}%
dk\frac{\,k}{\kappa}\left\{  \zeta_{n}^{2}\left(  \frac{r^{2}}{r^{2}%
-e^{2\kappa a}}+\frac{r^{\prime}{}^{2}}{r^{\prime}{}^{2}-e^{2\kappa a}%
}\right)  +\right. \\
\left.  +k^{2}\left(  \frac{r}{1-r^{2}e^{-2\kappa a}}+\frac{r^{\prime}%
}{1-r^{\prime}{}^{2}e^{-2\kappa a}}\right)  e^{-\kappa a}\cosh\left[
\kappa(2z-a)\right]  \right\}   \, ,
\end{multline}
where $\zeta_n = 2\pi n/\beta$. The prime on the sum 
is a reminder to count the $n=0$ term with half weight, and
$\beta=1/kT$. This expression (which vanishes \ as $a\rightarrow\infty$)
represents the Casimir contribution to the energy density, or the difference
between the energy density at finite temperature with the dielectric walls
present and not. It does not include the energy density of a thermal bath
without the walls present. To get the latter energy density, we
can start with the full expression for $U$, which
includes the empty space vacuum divergent term \cite{Energy}. At zero
temperature, this term can be written as an integral over real frequencies:%

\begin{equation}
U_{ES}=-\frac{i}{2\pi^{2}}\int_{0}^{\infty}d\omega\int_{0}^{\infty}%
dk\,k\frac{\omega^{2}}{\kappa},\; \label{Ues}%
\end{equation}
where $\kappa$ is defined as $\kappa^{2}=k^{2}-\omega^{2}$. For finite
temperatures, this term is modified by inserting a factor $\left[
1+2/\left(  e^{\beta\omega}-1\right)  \right]$ to account for the thermal
energy. This factor reflects the fact that at zero temperature, each mode has
an energy of $\frac{1}{2}\omega$; at finite temperature, there is an
additional thermal energy of $1/\left(  e^{\beta\omega}-1\right)  $. After
removing the divergent term, the result is the familiar result for the
 energy density of blackbody radiation:

\begin{equation}
\triangle U_{ES}=\frac{\pi^{2}}{15\beta^{4}}.
\end{equation}
The energy density in the vacuum region at finite temperature is then%
\begin{equation}
U(T)=U_{C}+\triangle U_{ES}.
\end{equation}

\bigskip%
\begin{figure}
[ptb]
\begin{center}
\includegraphics[
trim=0.000000in 0.000000in -0.908808in 0.000000in,
height=2.7605in,
width=4.0145in
]%
{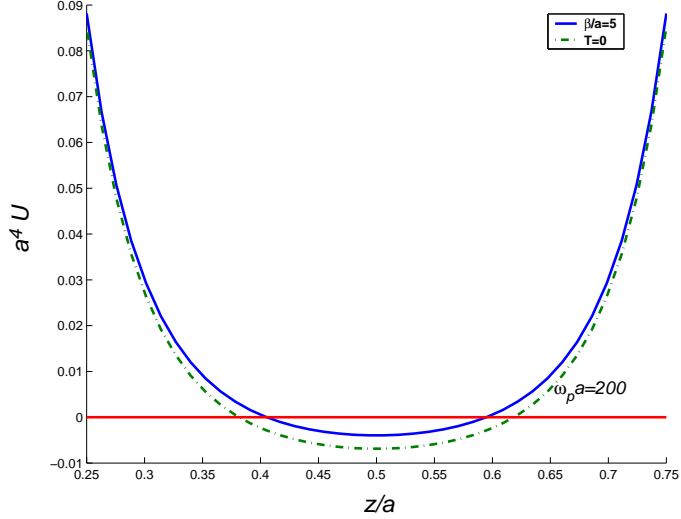}%
\caption{The solid curve represents the energy density at finite temperature
corresponding to $\beta/a=5$, as compared to the energy density at zero
temperature (dash-dot line). As expected, the local energy density increases
with temperature, and negative energy density is only possible if the
temperature is sufficiently low. }
\label{fig:Ub}
\end{center}
\end{figure}

Figure~\ref{fig:Ub} shows 
two graphs representing energy density at zero and finite
temperature corresponding to $\beta/a=5$. As expected, the local energy
density increases with temperature, and negative energy density is only
possible if the temperature is sufficiently low. On the other hand, the zero
temperature results are a good approximation so long as $\beta\gg a$. For
systems at room temperature, this increase in the energy density is still very
small when the separations between the walls are of the order of few
micrometers. More generally, one can ignore thermal effects at distances small
compared to $1/(kT)$. In this case, it is still possible to achieve negative
energy density in the central region. At room temperature, for example,
$\beta \approx 8 \, \mu{\rm m}$. Thus, $\beta/a=5$ corresponds to $a = 1.6\,\mu{\rm m}$,
which is in the range of separations for which Casimir force experiments have
been performed. 

\begin{figure}
[ptb]
\begin{center}
\includegraphics[
trim=0.000000in 0.000000in -1.567125in 0.000000in,
height=2.7605in,
width=4.0145in
]%
{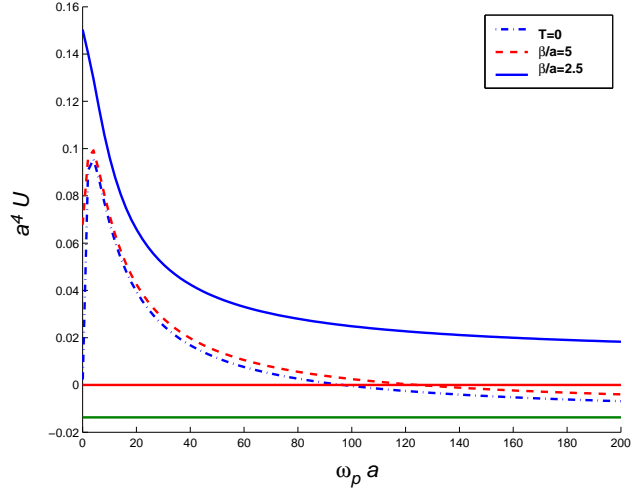}%
\caption{The figure represents the energy density at the center of the vacuum
region as a function of $\omega_{p}a$ for various temperatures including zero
temperature (dash-dot line). As the temperature grows (and $\beta$ decreases),
the value of $\omega_{p}a$ 
for which the energy density becomes negative shifts towards
larger values.}
\label{fig:Ucenter}
\end{center}
\end{figure}

The energy density at the center of the vacuum region is shown in 
Fig.~\ref{fig:Ucenter} as a function  of $\omega_{p}a$ for various 
temperatures. As the temperature increases, the region of negative energy
density shrinks, and when $\beta \lesssim 2.6\, a$, the energy density is
positive everywhere.

\section{The Transverse and Longitudinal Pressure}
\label{sec:press}

First consider the longitudinal pressure, $p_z =   T_{zz} $. 
The conservation law,
$\partial^\mu \,  T_{\mu\nu} = 0$, with $\nu = z$, and the fact that
$  T_{\mu\nu} $ is diagonal, implies that 
$\partial^z \,  T_{zz} = 0$. Thus
$  T_{zz} $ is constant. From the relation 
$T_{zz} = T_{00} - E_z^2 -B_z^2$,
we find that, at zero temperature,

\begin{equation}
 p_z = T_{zz} =
\frac{1}{2\pi^{2}}\int_{0}^{\infty}d\zeta\int_{0}^{\infty}%
dk\;k\, \kappa\left(  \frac{r^{2}}{r^{2}-e^{2\kappa a}}+\frac{r^{\prime}{}^{2}%
}{r^{\prime}{}^{2}-e^{2\kappa a}}\right)  \label{eq:Tzz}%
\end{equation}

\begin{figure}
[ptb]
\begin{center}
\includegraphics[
trim=0.000000in 0.000000in -1.019322in 0.000000in,
height=2.7605in,
width=4.0145in
]%
{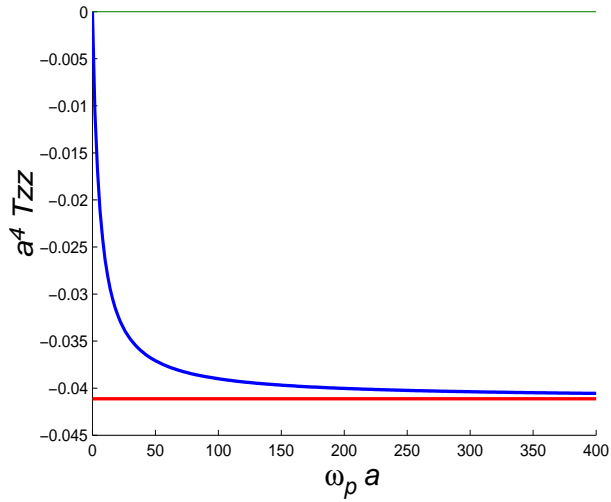}%
\caption{The graph represents the pressure $T_{zz}$ as a
function of $\omega_{p}a$ at zero temperature for all
 $0<z<1$. The horizontal line corresponds to 
the perfectly reflecting wall. }%
\label{fig:TzzCenter}%
\end{center}
\end{figure}

The plot of this function is shown in Fig.~\ref{fig:TzzCenter} as a
function of $\omega_{p}a$ at any value of z in the vacuum region. 
The horizontal
line corresponds to the perfectly reflecting wall, 
$p_z =-\pi^{2}/\left( 240a^{4}\right) $. Note that $p_z$ is the force per
unit area on one half-space due to the other, and agrees with the result
of the Lifshitz theory~\cite{Schwinger,Lif}. 
The magnitude of the force is maximum in the perfectly
conducting limit.  

The transverse pressure, 
$p_x =  T_{xx}  = p_y =  T_{yy} $ is a 
nontrivial function of $z$,
and is given at zero temperature by
\bigskip%
\begin{eqnarray}
 T_{xx} =-\frac{1}{4\pi^{2}}\int_{0}^{\infty}d\zeta\int_{0}^{\infty
}dk\frac{\,k^{3}}{\kappa}\left\{  \left(  \frac{r^{2}}{r^{2}-e^{2\kappa a}%
}+\frac{r^{\prime}{}^{2}}{r^{\prime}{}^{2}-e^{2\kappa a}}\right)  -\right. 
   \nonumber   \\
\left.  -\left(  \frac{r}{1-r^{2}e^{-2\kappa a}}+\frac{r^{\prime}}%
{1-r^{\prime}{}^{2}e^{-2\kappa a}}\right)  e^{-\kappa a}\cosh\left[
\kappa(2z-a)\right]  \right\}  \, , \label{eq:px}
\end{eqnarray}

\begin{figure}
[ptb]
\begin{center}
\includegraphics[
trim=0.000000in 0.000000in -1.019322in 0.000000in,
height=2.7605in,
width=4.0145in
]%
{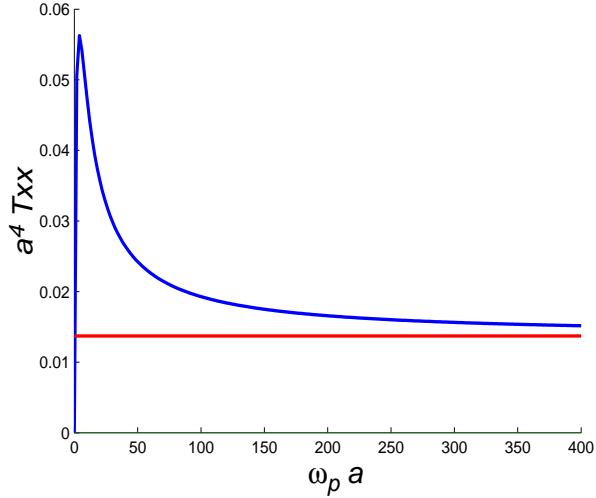}%
\caption{The graph represents the pressure $T_{xx}$ at zero temperature as a
function of $\omega_{p}a$ at $z=0.5 a$. The horizontal line corresponds to the 
perfectly reflecting wall.}%
\label{fig:TxxCenter}%
\end{center}
\end{figure}

The plot of this function is shown in Fig.~\ref{fig:TxxCenter} as a
function of $\omega_{p}a$ at the center of the vacuum region. The horizontal
line corresponds to the perfectly reflecting wall.

If we compare the position dependent term in $U$, Eq.~(\ref{eq:U}), with 
that in 
$T_{xx}$, Eq.(\ref{eq:px}), we see that they differ by a factor of two. 
These terms are dominant near a wall, so 
the asymptotic form of $T_{xx}$ near the boundary $z=0$ is one half the 
corresponding expression
for $U$ in this limit, which was found in Ref.~\cite{Energy}: 
\begin{equation}  
T_{xx} \sim \frac{1}{2}\, U \sim 
\frac{\sqrt{2}\omega _{p}}{128\pi }\frac{1}{z^{3}} \,, 
                     \qquad {\rm as}\quad z \rightarrow 0 \,.
  \label{TxxLim}  
\end{equation}  
Again, this singular behavior arises despite the inclusion of
dispersion in our treatment, and can be viewed as due to the assumption
of a sharp boundary at $z = 0$. A crucial point is that the reflection 
coefficients vanish as $\omega^{-2}$ as $\omega \rightarrow \infty$.
Pfenning~\cite{Pfenning} has studied scalar models in which these coefficients
vanish more rapidly at high frequency, and obtained a finite stress tensor
at the boundary.

\section{\bigskip The Null Energy Condition}

Now we turn to the null energy condition for rays travelling parallel 
to the walls.
Combining Eqs.~(\ref{eq:U})\textbf{\ (}$T_{00}=U$\textbf{) }and 
(\ref{eq:px}), and changing variables using $\zeta = u\, t$ and 
$k = u\ \sqrt{1-t^2}$, we have

\begin{multline}
T_{00}+T_{xx}=\frac{1}{4\pi^{2}}\int_{0}^{\infty}du\,u^{3}\int_{0}%
^{1}dt\left\{  \left(  3t^{2}-1\right)  \left(  \frac{r^{2}}{r^{2}-e^{2ua}%
}+\frac{r^{\prime}{}^{2}}{r^{\prime}{}^{2}-e^{2ua}}\right)  \right. \\
\left.  +3\left(  1-t^{2}\right)  \left[  \frac{r}{1-r^{2}e^{-2ua}}%
+\frac{r^{\prime}}{1-r^{\prime}{}^{2}e^{-2ua}}\right]  e^{-ua}\cosh\left[
u\left(  2z-a\right)  \right]  \right\} \, . \label{Null}
\end{multline}

We see that in
case of $r\rightarrow-1$ and $r^{\prime}\rightarrow1$, or perfectly reflecting
walls, $T_{00}+T_{xx}=0$, so the null energy condition is marginally
satisfied. We now wish to show that in all other cases, $T_{00}+T_{xx}>0$.
First consider the terms proportional to $3t^{2}-1$. Of these, the term
proportional to $r^2$ vanishes  because $r$ is
independent of $t$ and $\int_{0}^{1}dt\;\left(  3t^{2}-1\right)  =0$. 
Next consider the term proportional to  $3t^{2}-1$ and to $r^{\prime}{}^{2}$.
 Its contribution to the integral on
$t$  can be written, using partial integration, as
\begin{equation}
\int_{0}^{1}dt\;\left(  3t^{2}-1\right)  \frac{r^{\prime}{}^{2}}{r^{\prime}%
{}^{2}-e^{2ua}}=-e^{-2ua}\int_{0}^{1}dt\;t\left(  1-t^{2}\right)
\frac{2r^{\prime}{}}{\left(  1-r^{\prime}{}^{2}e^{-2ua}\right)  ^{2}}%
\frac{dr^{\prime}}{dt}, \label{tInt}%
\end{equation}
where in terms of the new coordinates,
\begin{equation}
r^{\prime}=
\frac{\,u^{2}t^{2}+\omega_{p}^{2}-ut^{2}\sqrt{u^{2}+\omega_{p}^{2}}}
{\,u^{2}t^{2}+\omega_{p}^{2}+ut^{2}\sqrt{u^{2}+\omega_{p}^{2}}} \,.
\end{equation}
Its derivative,
\begin{equation}
dr^{\prime}/dt=- \frac{4u\omega_{p}^{2}t\sqrt{u^{2}+\omega_{p}^{2}}}
{\left[
ut^{2}(\,u+\sqrt{u^{2}+\omega_{p}^{2}})+\omega_{p}^{2}\right]^{2}}\,,
\end{equation}
is negative. Thus,
\begin{equation}
\int_{0}^{1}dt\;\left(  3t^{2}-1\right)  \frac{r^{\prime}{}^{2}}{r^{\prime}%
{}^{2}-e^{2ua}} > 0 \,.
\end{equation} 
Now consider the term in Eq.~(\ref{Null}) proportional to $1-t^2$. 
We can write
\begin{equation}
r = \frac{u -\sqrt{u^2+\omega_p^2}}{u +\sqrt{u^2+\omega_p^2}}\, ,
\end{equation}
from which we see that $r < 0$. We can also show that
\begin{equation}
r'-|r| = r' +r =
\frac{2 \omega_p^2 (1-t^2) u}
{ \sqrt{u^2+\omega_p^2}\,(2t^2u^2+\omega_p^2) +2t^2u^3 +u\, \omega_p^2 (t^2+1)}
\geq 0 \, ,
\end{equation}
for $0\leq t \leq 1$,  from which it follows that  $r' \geq |r| \geq 0$.
This implies that the $1-t^2$ term in Eq.~(\ref{Null}) is non-negative.
Thus $T_{00}+T_{xx} \geq 0$, and the null energy condition is satisfied
by a finite margin, except for the limiting case of a perfect conductor. 
This implies that the gravitational effect on light rays
moving 
parallel to the plates is to cause focusing. Even though the energy density
can be negative, its effect is more than cancelled by the positive pressure.

\bigskip Now we inspect the effect of the finite temperature on the 
null energy condition. We expect that finite temperature would make the
null energy condition satisfied by a wider margin. So, we write
Eq.~(\ref{Null})\ as Fourier series as above in Eq.~(\ref{Uct}):%

\begin{multline}
T_{00}+T_{xx}=\frac{1}{2\pi\beta}\sum_{n=0}^{\infty}{}^{\prime}\int
_{0}^{\infty}dk\frac{\,k}{\kappa}\left\{  \left(  k^{2}-2\zeta_{n}^{2}\right)
\left(  \frac{r^{2}}{e^{2\kappa a}-r^{2}}+\frac{r^{\prime}{}^{2}}{e^{2\kappa
a}-r^{\prime}{}^{2}}\right)  -\right. \\
\left.  -3k^{2}\left(  \frac{r}{r^{2}e^{-2\kappa a}-1}+\frac{r^{\prime}%
}{r^{\prime}{}^{2}e^{-2\kappa a}-1}\right)  e^{-\kappa a}\cosh\left[
\kappa(2z-a)\right]  \right\}  +\frac{4\pi^{2}}{45\beta^{4}}.
\end{multline}

Evaluated at $z=0.5a$, where it has a minimum, this becomes:%

\begin{multline}
\label{NullTcenter}\left(  T_{00}+T_{xx}\right)  _{z=0.5a}=
\frac{1}{2\pi\beta}\sum_{n=0}^{\infty
}{}^{\prime}\int_{\zeta_{n}}^{\infty}d\kappa\left\{  \left(  \kappa^{2}%
-3\zeta_{n}^{2}\right)  \left(  \frac{r^{2}}{e^{2\kappa a}-r^{2}}%
+\frac{r^{\prime}{}^{2}}{e^{2\kappa a}-r^{\prime}{}^{2}}\right)  -\right. \\
\left.  -3\left(  \kappa^{2}-\zeta_{n}^{2}\right)  \left(  \frac{r}%
{r^{2}e^{-2\kappa a}-1}+\frac{r^{\prime}}{r^{\prime}{}^{2}e^{-2\kappa a}%
-1}\right)  e^{-\kappa a}\right\}  +\frac{4\pi^{2}}{45\beta^{4}}.
\end{multline}

\begin{figure}
[ptb]
\begin{center}
\includegraphics[
trim=0.000000in 0.000000in -1.019322in 0.000000in,
height=2.7605in,
width=4.0145in
]%
{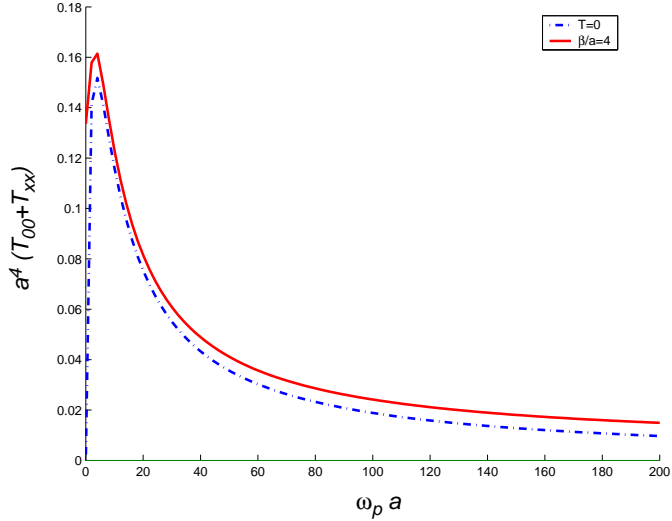}%
\caption{The solid curve represents the energy density plus pressure as a
function of $\omega_{p}a$ at $z=0.5a$ and at finite temperature corresponding
to $\beta/a=5$. The dash-dot line represents the same function at zero
temperature. The graph indicates that finite temperature makes the 
null energy condition in the transverse direction 
satisfied by a wider margin, as expected. }%
\label{nullb4}%
\end{center}
\end{figure}

\bigskip A change of variables $k\longrightarrow\kappa$ has been made. A plot
of Eq.~(\ref{NullTcenter}) as a function of $\omega_{p}a$ is shown in 
Fig.~\ref{nullb4}. The quantity $ T_{00}+T_{xx}$ is always positive and
increases with increasing temperature, so the null energy condition is always
satisfied for transverse rays.

\bigskip%

\bigskip Now we examine the null energy condition for light rays perpendicular
to the walls.  Adding Eqs.~(\ref{eq:Tzz}) and (\ref{eq:U}) yields, at 
zero temperature

\begin{multline}
\label{UTzz}T_{00}+T_{zz}=\frac{1}{2\pi^{2}}\int_{0}^{\infty}du\,u^{3}\int_{0}%
^{1}dt\left\{  \left(  1+t^{2}\right)  \left(  \frac{r^{2}}{r^{2}-e^{2ua}%
}+\frac{r^{\prime}{}^{2}}{r^{\prime}{}^{2}-e^{2ua}}\right)  \right.  \\
\left.  +\left(  1-t^{2}\right)  \left[  \frac{r}{1-r^{2}e^{-2ua}}%
+\frac{r^{\prime}}{1-r^{\prime}{}^{2}e^{-2ua}}\right]  e^{-ua}\cosh\left[
u\left(  2z-a\right)  \right]  \right\}  .
\end{multline}

\begin{figure}
[ptb]
\begin{center}
\includegraphics[
trim=0.000000in 0.000000in -0.945700in 0.000000in,
height=2.7605in,
width=4.0145in
]%
{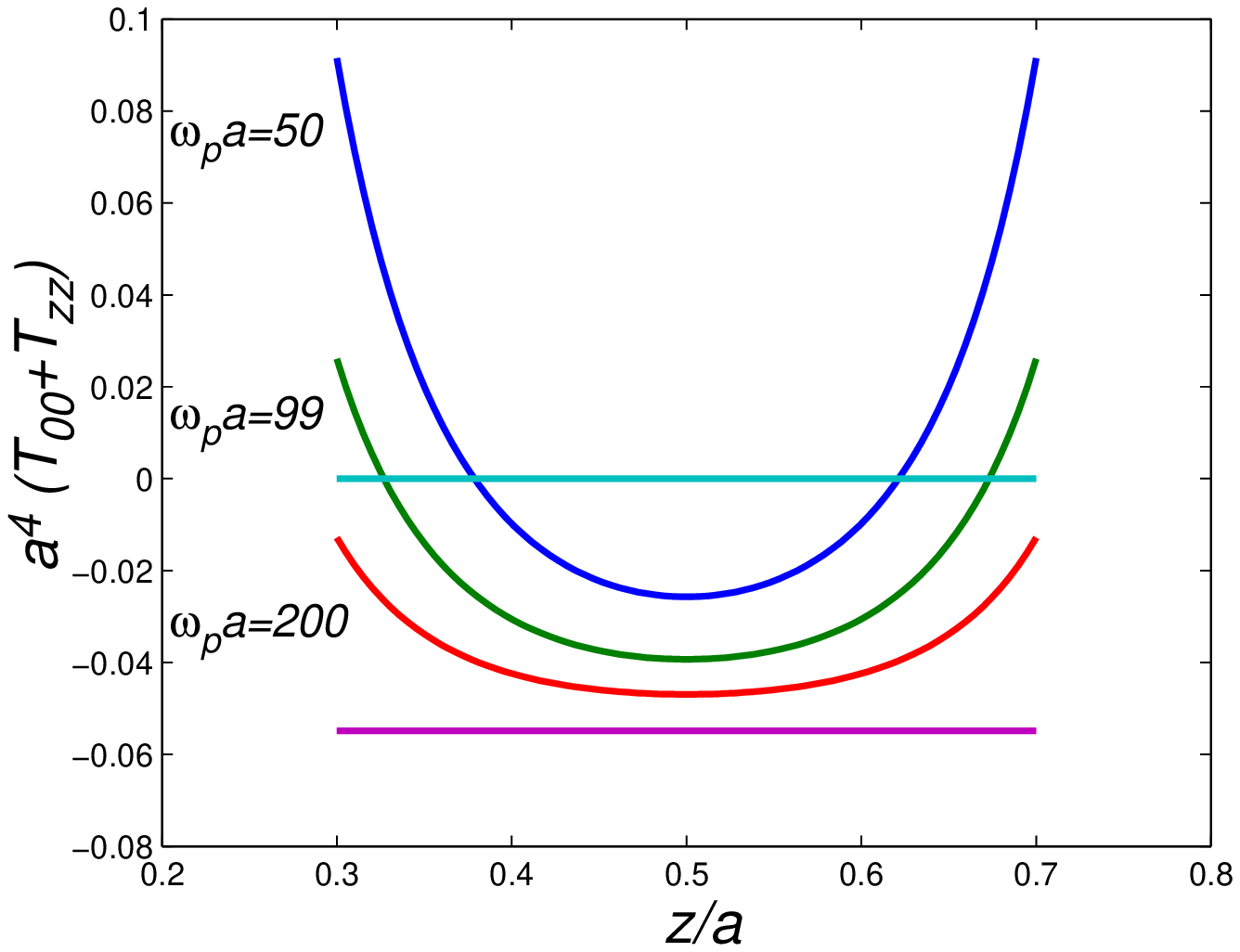}%
\caption{A plot of $T_{00}+T_{zz}$ as a function of position is shown for three
values of $\omega_{p}a$ at zero temperature. 
The bottom solid line corresponds to the case of
perfectly reflecting walls, for which the null energy condition in $z$
direction is violated. The figure shows that in the case of dielectric walls, 
the null energy condition can be violated locally in this direction.}%
\label{utzz3}
\end{center}
\end{figure}

A plot of Eq.~(\ref{UTzz}) as a function of position is shown in 
Fig.~\ref{utzz3} for
three values of $\omega_{p}a$. The bottom solid line corresponds to the case
of perfectly reflecting walls, $T_{00}+T_{zz}=-\pi^{2}/\left(  180a^{4}%
\right)  $, for which the null energy condition in the $z$ direction is
violated. The figure shows that in the case of dielectric walls, 
the null energy
condition can be violated locally in this direction, but only over a
 restricted interval in the internal region. The
average null energy condition integral, Eq.~(\ref{eq:anec}), will
acquire a net positive contribution even before a null ray reaches a 
boundary. Finite temperature will further restrict the region
where the null energy condition can be violated.

\section{Summary}

In this paper, we have discussed the effects of finite reflectivity of
the walls and of finite temperature on the Casimir energy density and 
pressures. In particular, we have been interested in when the weak
energy condition and the null energy condition can be violated. We find
that the weak energy condition and the null energy condition for rays
not parallel to the plates can still be violated, but with more difficulty 
than in the case of perfectly reflecting plates. Furthermore, these 
violations are now confined to a localized central region finitely
removed from the boundaries. These regions decrease in size as the 
temperature increases, and eventually vanish for $\beta \ll a$.
 The energy density and transverse pressure
diverge positively as the boundaries are approached, further limiting
the region of possible energy condition violation. The null energy 
condition for rays parallel to the boundary, which is marginally satisfied
for the perfectly reflecting case at zero temperature, is satisfied by
a finite margin with either finite reflectivity or finite temperature.

\newpage

\centerline{\bf Acknowledgments}
\vspace{0.5cm}
This work was partially supported by NSF grant PHY-0244898.

\end{document}